\documentclass[final,5p,times,twocolumn]{elsarticle}

\usepackage{amssymb}

\usepackage{amsmath}
\usepackage{subcaption}
\usepackage{adjustbox}
\usepackage{multirow}
\usepackage[hyphens]{url}
\usepackage{xcolor}
\usepackage{graphics}

\journal{Technological Forecasting and Social Change}

\begin{document}

\begin{frontmatter}



\title{Modeling Technological Deployment and Renewal:\\Monotonic vs. Oscillating Industrial Dynamics}


\author[inst1]{Joseph Le Bihan\corref{correspondingauthor}}
\ead{joseph.le-bihan@etu.u-paris.fr}
\cortext[correspondingauthor]{Corresponding author. : J. Le Bihan, Université Paric Cité, Bâtiment Condorcet, 10 rue Alice Domon et Léonie Duquet, 75013, Paris, France}

\affiliation[inst1]{organization={Université Paris Cité, CNRS, LIED UMR 8236},
            city={Paris},
            postcode={F-75006},
            country={France}}

\author[inst1]{Thomas Lapi}
\ead{thomas.lapi@etu.u-paris.fr}
\author[inst1]{José Halloy}
\ead{jose.halloy@u-paris.fr}

\begin{abstract}

This study proposes a new model based on a classic S-curve that describes deployment and stabilization at maximum capacity. In addition, the model extends to the post-growth plateau, where technological capacity is renewed according to the distribution of equipment lifespans. We obtain two qualitatively different results. In the case of “fast” deployment, characterized by a short deployment time in relation to the average equipment lifetime, production is subject to significant oscillations. In the case of “slow” deployment, production increases monotonically until it reaches a renewal plateau. These results are counterintuitively validated by two case studies: nuclear power plants as a fast deployment and smartphones as a slow deployment. These results are important for long-term industrial planning, as they enable us to anticipate future business cycles. Our study demonstrates that business cycles can originate endogenously from industrial dynamics of installation and renewal, contrasting with traditional views attributing fluctuations to exogenous macroeconomic factors. These endogenous cycles interact with broader trends, potentially being modulated, amplified, or attenuated by macroeconomic conditions. This dynamic of deployment and renewal is relevant for long-life infrastructure technologies, such as those supporting the renewable energy sector and has major policy implications for industry players.
\end{abstract}


\end{frontmatter}
\section{Introduction}
\label{sec:intro}

The S-curve (also known as the sigmoid or logistic curve) is a frequently used model for technology or innovation diffusion \cite{griliches1957,rogers_diffusion_1983}. This curve typically exhibits three distinct phases: an initial period of slow growth, followed by a phase of rapid expansion, and finally, a stage of saturation with minimal further increase. This type of dynamic has been observed for a wide range of \textit{technological equipment} including transportation systems \cite{grubler1990}, energy generation methods \cite{iiasa1024,rao_review_2010,cherp_national_2021} and various consumer goods \cite{bass1969,peres_innovation_2010}.

Researchers have used mathematical modelings of varying complexity and underlying assumptions to model and understand these dynamics \cite{meade2006modelling,guidolin_innovation_2023}. These models focus primarily on the drivers of the rapid expansion phase, and are used to forecast the level of adoption of a technology, or to anticipate the dominant technologies of the future \cite{kucharavy_application_2011}.

While the initial growth and rapid expansion phases of technological diffusion have been widely studied, the dynamics of the post-growth steady state remain comparatively understudied. This lack often stems from the prevailing assumption that new technologies will displace existing ones, triggering a new S-shaped diffusion curve. By ``new technology", we mean here a technological discontinuity, a revolutionary breakthrough that entails, among other things, changes in the technical skills, resources and processes required to design and produce the item, as well as physical changes in the item itself \cite{ehrnberg_definition_1995}. The evolution of music storage and lighting technologies are examples of successive technological discontinuities in line with the previous assumption.

The progression from vinyl records to magnetic tapes, CDs, and ultimately digital formats in music storage represents a series of breakthrough innovations. Each format constitutes a distinct technological platform with novel functionalities, operating principles, and material composition. Similarly, the transition from incandescent lamps to fluorescent lamps and, more recently, to light-emitting diodes (LEDs) showcases a comparable pattern. In both these cases, the classic S-curve is not readily applicable up to its characteristic plateau phase. The emergence of a new technology disrupts the existing market, fostering the development of entirely new supply chains alongside its own growth trajectory.

On the other hand, the evolution of a technology can also be continuous. A more or less lengthy period of incremental change generally precedes a revolutionary breakthrough \cite{anderson_technological_1990}. During this period of incremental evolution, the dominant design of a technology undergoes advances or refinements within the same fundamental technological framework. These technological advancements do not disrupt the material composition, production processes or operating principle. The Apple iPhone, for instance, exemplifies this category. Despite significant internal advancements since its 2007 inception, the core functionality and form factor remain largely unchanged. Similarly, transistors, a cornerstone of modern electronics, have undergone substantial performance enhancements while retaining silicon or germanium as the primary material.

For technologies exhibiting a long incremental evolution period, the post-growth phase holds particular significance. Unlike revolutionary breakthrough that introduce entirely new categories of devices, these technologies often experience a plateau in the total number of units in service. This can be observed in infrastructure projects like the total railroad track length or consumer electronics like mobile phone saturation. Notably, innovation in these cases often occurs at the end-of-life (EoL) stage. As existing equipment reaches its functional obsolescence, it is replaced by newer, potentially more efficient iterations of the same core technology.

For these technologies, a post-growth plateau is experienced, and the total number of units in service is maintained thanks to a renewal dynamic. Current technology diffusion models, which focus mainly on the initial growth phase, do not address this phase, where unit production is no longer driven by technology deployment, but by renewal. Here, ``renewal" means the sustained provision of a service by a technology undergoing continuous, non-disruptive evolution.

For this ``renewal" phase, the object of interest is no longer the number of units in service (which is maintained at the saturation value), but the production of units to replace those reaching the end of their service life. This concept of renewal production to maintain a set of active technological equipment on a global scale remains understudied in the existing literature. While pioneering works in the 1930s explored this domain \cite{lotka1933,lotka1939,preinreich1939theory}, the focus subsequently shifted towards more abstract and generalizable models, culminating in the development of Renewal Theory – a powerful yet highly theoretical framework with limited applicability in real-world industrial settings. 

Despite some attempts in the Marketing literature \cite{bayus1989developing}, no comprehensive model has been produced with the aim of linking technology deployment and renewal by modeling the production of technological equipment on a consistent timescale. This study bridges this gap by proposing a technology deployment model, based on the literature cited above, coupled with a simple renewal mechanism to model equipment production over the long term. 

\section*{Positioning the problem through the examples of smartphones and nuclear power plants}
\label{sec:ComparisonIntro}

The aim of this short section is to highlight the existence of renewal phases for certain technologies, where the number of equipment in service varies little. And to justify the interest of a model by the shape of the dynamics observed. 

\begin{figure*}[h]
    \centering
    \includegraphics[width = \textwidth]{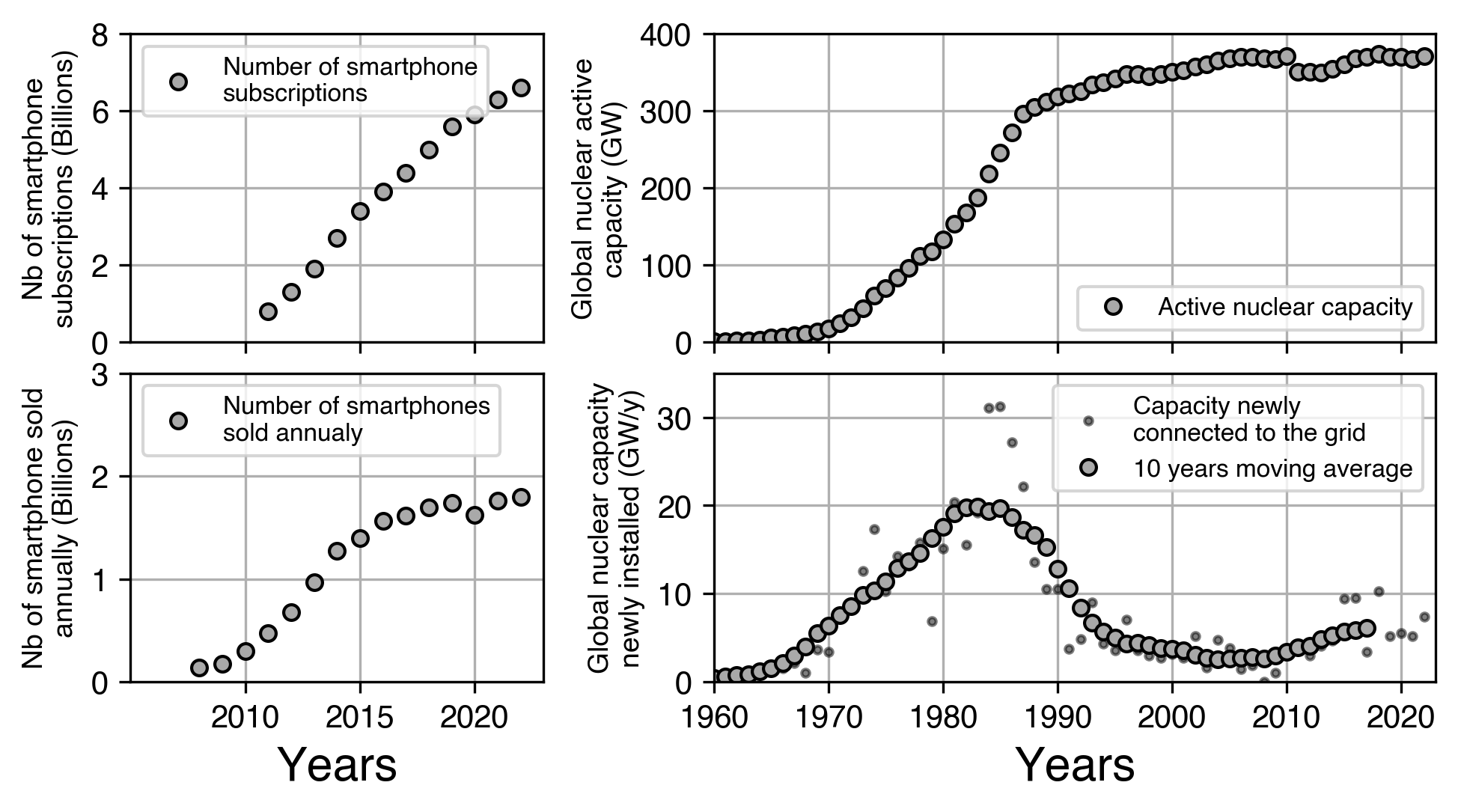}
    \caption{Comparison between Smartphones (left) and Nuclear Plants (right) deployment dynamics. Upper panels display the technology adoption or deployment, through number of active smartphone subscriptions and active nuclear capacity. Lower panels display the ``equipment production'' supporting the deployment, through annual smartphone sales and nuclear capacity novel connections to the grid.}
    \label{fig:IntroComparison}

\end{figure*}

Figure \ref{fig:IntroComparison} shows the deployment of Smartphone and Nuclear Power Plant technologies, and the dynamics of their technological equipment production. The renewal phase is clear for nuclear power, with little change in active capacity since the 1990s (upper right panel). For smartphones, growth in the number of subscriptions, or ``active'' smartphones, seems to be slowing down (upper left panel), pointing to saturation. Furthermore, this saturation has already been observed for cellular mobile subscriptions \cite{worldbank2022} and therefore seems a reasonable assumption for smartphones.

With adapted time scales, the deployment dynamics (upper panels) appear similar. However, the equipment production curves (lower panels) appear to be qualitatively different: for smartphones, production no longer seem to vary significantly, whereas for power plants, a peak is observed. 

Such differences in production dynamics necessarily have important effects on the underlying industry. It is therefore essential to develop a framework for understanding the drivers of these dynamics, and to consider both potential effects and areas for action.

\section*{Objectives}
\label{sec:objectives}

This study investigates the gap in existing literature concerning the combined analysis of deployment and renewal phases of a technology global in-use equipment fleet. We propose a novel, parsimonious model to describe the evolution of equipment production, encompassing deployment and subsequent renewal periods. The aim of this model is to provide comprehensible answers to the questions raised by Figure \ref{fig:IntroComparison}, and thus to explain the possibility of qualitatively different behavior for deployments.

The focus is on the production constraints that arise when the deployment of a technology saturates or reaches a predetermined capacity, and when this level is maintained over a substantial period of time through technological equipment renewal. By analyzing the production profiles required to maintain a constant level of active technology after deployment, this study reveals the existence of two distinct post-growth industrial dynamics.

Within this framework, ``active technology" or ``capacity" refers to the operational equipment count supporting the technology (upper panels of Figure \ref{fig:IntroComparison}). This quantity is characterized by an S-shaped curve, reflecting growth (deployment phase) followed by a stable state (renewal phase).

The central research question deals with the potential production dynamics that harmonize with the aforementioned capacity curve, while factoring in the replacement of equipment reaching its EoL stage. The study aims to identify the production constraints associated with each possible dynamic.

This study is organized as follows.

\begin{itemize}
    \item {\bf Section~\ref{sec:model}:}  This section covers the modeling framework used in the analysis.
    \item {\bf Section~\ref{sec:AnalysisKeyParam}:}  Here, we present the mathematical demonstrations and main solutions of the model that describe two possible types of dynamics, with and without damped oscillations. 
    
    \item {\bf Section~\ref{sec:CaseStudies}:} In this section, we present the application of the model to the two examples mentioned: the nuclear power plant industry and the smartphone industry; and discuss its validity.
    
    \item {\bf Section~\ref{sec:Discussion}:} This section presents a discussion of the various results and limitations of the model, and considers possible future extensions and applications.
    
    \item {\bf Section~\ref{sec:IndustrialConsequences}:} This section summarizes the model results and sets out the main implications for industrial strategies.
    
\end{itemize}

\section{Defining a generic mathematical model for technology deployment and renewal}
\label{sec:model}

We model here the behavior of technology equipment production from the technology deployment to its renewal on a relatively long term. The production will be derive from two parameters characterizing the dynamics : the \textit{S-shaped} curve describing the growth of the number of equipment in-use and the EoL distribution of an equipment. 

\subsection{Modeling technology deployment}

The study of technological deployment or diffusion has a vast history, dating back to the early 20th century. Pioneering work by Lotka (1926) used the logistic curve to model the growth of the American railway system, laying the groundwork for subsequent research \cite{lotka1926}. Since then, a vast body of literature has employed S-shaped curves to describe the diffusion patterns of new technologies \cite{kuznets1930,griliches1957,mansfield1961,bass1969,fisher1971,iiasa1024,grubler1990,grubler1999}.

In line with established research on technological diffusion, we will employ an S-shaped curve to model the deployment trajectory of the target technology. This approach aligns with the observation that the number of deployed technological units (or, when applicable, a directly related metric such as power capacity for energy technologies) over time (denoted by t) can be effectively captured by a logistic function:
\begin{equation}
\label{eq:capa}
Capacity(t) = \frac{K}{1+e^{-\left(t-t_{peak}\right)/\tau_{dep}}}    
\end{equation}

Our choice of the logistic curve is motivated by its parsimony and recognized usefulness in modeling technological diffusion. The simple parameterization framework provided by the logistic function makes it easy to adapt the dynamic analysis to other S-curves (Gompertz function, cumulative Gaussian distribution), and the results are not qualitatively different. The parameters are as follows:

\begin{itemize}
    \item $\mathbf{t_{peak}}$ - Peak production time: this parameter corresponds to the point at which deployment speed is at its maximum, i.e. when production is at its peak. It separates the acceleration phase of the deployment process from its deceleration phase. 
    \item $\mathbf{K}$ - Carrying capacity: this parameter denotes the saturation level or the target capacity of the technology. It represents the number equipment units to be renewed over the long term.
    \item $\boldsymbol{\tau}_{dep}$ - Characteristic deployment time: This parameter quantifies the rate or pace of the deployment process. To ensure mathematical tractability and facilitate clear calculations, we adopt the canonical form of the logistic function (as presented in Eq. (\ref{eq:capa})). Consequently, $\tau_{dep}$ is implicitly defined as half the time it takes for deployment to progress from 27\% to 73\% of the ultimate carrying capacity. It is noteworthy that defining $\tau_{dep}$ based on a different growth interval would solely affect the time scale and not fundamentally alter the underlying deployment dynamics.
\end{itemize}

To track this deployment of equipment in service, the instantaneous production of equipment, i.e. the production of equipment per unit of time, should be as follows :
\begin{equation} \label{eq:proddep}
    P_{dep}(t) = \frac{d}{dt}Capacity(t) = \frac{K}{\tau_{dep}} \times \frac{e^{-\left(t-t_{peak}\right)/\tau_{dep}}}{\left(1+e^{-\left(t-t_{peak}\right)/\tau_{dep}}\right)^2} 
\end{equation}

Technological equipment is produced to increase active capacity and follows the deployment profile given by the S-shaped curve.

\subsection{Renewal constraint}

The model presented so far assumes an idealized scenario in which deployed equipment remains operational for an indefinite period.
This idealized scenario, however, has only limited practical application, as technological equipment inevitably depreciates and must be replaced over time.

If this problem has little influence when we focus solely on initial purchases (as studied by \cite{bass1969}) or when technological substitution occurs before the equipment  lifespan (as studied by \cite{fisher1971,iiasa1024,grubler1990}), it gives rise to a non-negligible production when we analyze the dynamics over a longer period of time, particularly when we reach the post-growth plateau.

While traditional technology deployment studies often neglect the critical issue of equipment renewal or industrial replacement, this topic has received attention within other fields. Early considerations of industrial replacement emerged from the domain of actuarial science, framed as an actuarial problem \cite{herbelot1909}. As Alfred Lotka (1939) succinctly described it, the problem centered on ``the number of annual accessions required to maintain a body of N policyholders constant, as members drop out by death." \cite{lotka1939}. Building upon these actuarial foundations, Lotka himself extended the concept to industrial replacement \cite{lotka1933,lotka1939}. This line of inquiry laid the groundwork for the development of renewal theory, a now-established branch of probability theory \cite{feller1957,smith1958}. The fundamental contributions of Feller (1941, 1949) were key to the development of this field \cite{feller1941,feller1949}.

The field of renewal theory has evolved considerably since its first applications in industrial contexts. The emphasis has been placed on a more theoretical framework, favoring abstract mathematical questions and generalizable results. Although our current work focuses on practical results for specific industrial quantities, we will maintain links with renewal theory wherever possible and without introducing excessive complexity.

To incorporate equipment production necessitated by EoL replacements into Eq. (\ref{eq:proddep}), we define the EoL distribution of an equipment : $p_{EoL}$. The probability that an equipment reach EoL $\theta$ years after its production is then $p_{EoL}(\theta)$. Furthermore, we assume that the number of equipment units produced is sufficiently large to invoke the law of large number: over $P_{tot}(t-\theta)$ equipment units produced at $t-\theta$ a fraction $P_{tot}(t-\theta)\times p_{EoL}(\theta)$ reaches EoL at $t$.

This fraction reaching EoL must then be sum for all the possible lifespan. Replacement production at years $10$ is the sum of units produced at year $0$ reaching EoL after $10$ years, produced at year $1$ reaching EoL after $9$ years and so forth. This sum covers all possible lifespan values, that is to say $0\leq \theta \leq +\infty$. In practice $\theta$ never exceeds a certain maximum lifespan $\tau_{MAX}$ so the sum could run between $0$ and $\tau_{MAX}$, however setting this boundary at infinity simplifies the mathematical resolution. 

Eq. (\ref{eq:proddep}) is reformulated to account for equipment production necessitated by EoL replacements. The resulting equation is presented below:

\begin{equation} \label{eq:model}
P_{tot}(t) = P_{dep}(t) + \int_{0}^{\infty}P_{tot}(t-\theta)p_{EoL}(\theta)d\theta
\end{equation}

where:
\begin{itemize}
\item $P_{tot}(t)$: Represents the total number of equipment units produced at time t.

\item $P_{dep}(t)$: Represents the production of new equipment units for deployment purposes at time t.

\item $p_{EoL}(\theta)$: Represents the EoL distribution, the probability density function of an equipment unit reaching its EoL $\theta$ years after production.

\item $\int_{0}^{\infty}P_{tot}(t-\theta)p_{EoL}(\theta)d\theta$: Represents the integral term that incorporates the production required to replace equipment produced at different points in time and reaching their EoL at $t$.

\end{itemize}

We employ a parsimonious parameterization for the EoL distribution $p_{EoL}$ characterized by the following parameters :
\begin{itemize}
    \item $\boldsymbol{\tau}_{EoL}$ - the average equipment lifespan before reaching EoL
    \item $\mathbf{CV_{EoL}} = \sigma_{EoL}/\tau_{EoL}$ - the coefficient of variation, \textit{i.e.} the variance of the equipment lifespan divided by the average lifespan. 
\end{itemize}

Our forthcoming analysis will demonstrate that the specific choice of probability distribution function employed to model the EoL phenomenon has a minimal impact on the overall system dynamics. This finding holds true as long as the two key parameters, average lifespan ($\boldsymbol{\tau}_{EoL}$) and coefficient of variation ($\mathbf{CV_{EoL}}$) are defined.

\subsection{Disaggregating technological adoption through replacement waves}

Following the derivation of Eq. (\ref{eq:model}), we can decompose the overall equipment production trajectory into two key constituents: the initial deployment phase $P_{dep}$, and the subsequent series of equipment replacement waves. Each technological equipment produced can be classified according to its role in this deployment and renewal dynamic:

\begin{itemize}
  \item Initial deployment: This category includes equipment produced for the initial adoption of the technology in question. These units are often referred to as initial purchases.
  \item First replacement: This category represents equipment produced to replace existing units that have reached their designated \textbf{EoL} for the first time. These replacements are commonly known as \textbf{second purchases}.
  \item Subsequent replacements: This category encompasses equipment produced for \textbf{ongoing replacements} that extend beyond the initial EoL cycle. This includes units procured for \textbf{third purchases, fourth purchases}, and so on.
\end{itemize}

This disaggregation by replacement waves allows for a more nuanced understanding of the dynamics of technological adoption within the overall equipment production function.

The number of equipment units requiring first replacement at time $t$, denoted by $R_1(t)$, can be mathematically modeled using convolution. This concept captures the cumulative effect of equipment deployed at different times reaching their EoL at $t$.

Here, $R_1(t)$ is expressed as the convolution of the initial deployment function, $P_{dep}(t)$, with the EoL probability density function, $p_{EoL}(\theta)$. This convolution is denoted by the asterisk ($\ast$) and represents the integral transformation:

$$R_1(t) = P_{dep} \ast p_{Eol}(t) = \int_{0}^{+\infty}P_{dep}(t-\theta)p_{EoL}(\theta)d\theta$$

Following the same logic, $R_2(t)$ represents the number of equipment units produced during the first replacement wave ($R_1$) that reach their EoL at time $t$. This can be expressed as the convolution of $R_1(t)$ with $p_{EoL}(\theta)$. Using the commutative property of the convolution the latter can be expressed as the convolution of $P_{dep}(t)$ with the convolution of $p_{EoL}(\theta)$ with itself,  
denoted by an exponent two ($p_{EoL}^{(2)}(\theta)$).

\begin{equation} \label{eq1}
\begin{split}
R_2(t) & = \int_{0}^{+\infty}R_{1}(t-\theta)p_{EoL}(\theta)d\theta \\
&= \int_{0}^{+\infty}\left(\int_{0}^{+\infty}P_{dep}(t-\theta-\theta')p_{EoL}(\theta')d\theta'\right)p_{EoL}(\theta)d\theta \\
&= \int_{0}^{+\infty}P_{dep}(t-\Theta)\left(\int_{0}^{+\infty}p_{EoL}(\Theta-\theta)p_{EoL}(\theta)d\theta\right)d\Theta \\
&\textit{with $\Theta = \theta' + \theta$}\\
&= \int_{0}^{+\infty}P_{dep}(t-\Theta)p_{EoL}^{(2)}(\Theta)d\Theta\\
&= P_{dep} \ast p_{EoL}^{(2)}(t)\\
\end{split}
\end{equation}

Iterating this procedure $n^{\text{th}}$ times allows to model the number of equipment units requiring a $n^{\text{th}}$ replacement at time $t$, or $n^{\text{th}}$ replacement wave, under the simple form :

$$R_n(t) = P_{dep} \ast p_{Eol}^{(n)} (t)$$

where $p_{EoL}^{(n)} (t)$ represents the $n^{\text{th}}$ iterated convolution of the EoL probability density function with itself.

The total equipment production for replacement at time $t$, $P_{tot}(t)$, can be obtained by summing the contributions from the initial deployment, $P_{dep}(t)$, and all subsequent replacement waves, $R_n(t)$:

\begin{equation}
P_{tot}(t) = P_{dep}(t) + \sum_{n=1}^{+\infty} R_n(t)
\end{equation}

This framework using convolution allows for a comprehensive analysis of equipment replacement dynamics within the broader context of technological adoption and production planning.

\subsection{Renewal process interpretation of equipment production} \label{sec:renewalprocess}

Equation (\ref{eq:model}) can be reinterpreted within the framework of renewal processes. This alternative perspective offers valuable insights into the dynamics of equipment production.

By setting $K = 1$ in Eq. (\ref{eq:model}) (without altering the underlying dynamics due to the proportionality between production and capacity), we can interpret $P_{dep}(t)$ as the probability of the only equipment ($K = 1$) initial installation at time $t$. Consequently, $P_{tot}(t)$ can be interpreted as the probability of either the initial installation or a replacement event occurring at time $t$.

This renewal process interpretation allows us to leverage established results from renewal theory regarding the existence and convergence properties of equipment production. However, a complete characterization of this production function requires a more in-depth analysis, beyond the basic framework presented here, and is carried out in the following section.

\section{Analysis of key parameters influencing renewal dynamics}
\label{sec:AnalysisKeyParam}
This section focuses on dissecting the qualitative impact of various parameters on the production dynamics associated with technology deployment and renewal. We specifically investigate the influence of:\\
$\ast$ Deployment characteristic time ($\tau_{dep}$);\\
$\ast$ Average EoL time constant ($\tau_{EoL}$);\\
$\ast$ Coefficient of variation of EoL (CV$_{EoL}$).

It is noteworthy that the parameters representing total capacity $(K)$ and peak production time $(t_{peak})$ do not exhibit a qualitative influence on the production dynamics.

A first result on the asymptotic behavior of Eq. \ref{eq:model}, whose formal demonstration can be found in \cite{feller1941}, is the convergence of production to a steady-state value.
In concrete terms, \textbf{both in-use capacity}, through the nature of the S-curve, \textbf{and equipment production}, through the properties of equation 3, converge towards a \textbf{Renewal Steady State (RSS)}. Mathematically, this steady state is characterized by:

$$
\left\{
\begin{array}{lll}
    Capacity(t) = K\\
\\
    P_{tot}(t) = \frac{K}{\tau_{EoL}}\\
\end{array}
\right.
$$

Within the context of our analysis, this result aligns with intuition.
These values depend neither on the deployment speed nor on the variance of the EoL distribution. Not surprisingly, these parameters no longer have any influence as the initial deployment wave becomes increasingly distant.
To maintain a constant in-use capacity of $K$ equipment units with an average active lifespan of $\tau_{EoL}$ years, the RSS dictates a yearly replacement rate of $K/\tau_{EoL}$ equipment units.

\subsection{Characterizing transient dynamics} \label{subsec:TwoBehav}

While Renewal Steady Production (RSP) is independent of deployment speed and equipment EoL distribution (aside from its average value), these factors do influence the system \textbf{transient dynamics}. This section explores the \textbf{transition behavior} between the initial \textbf{deployment regime} and the eventual \textbf{renewal regime}.

\subsubsection{Two types of behaviors}
The deployment regime is characterized by total production, $P_{tot}(t)$, being approximately equal to the deployment production, $P_{dep}(t)$. This signifies a period where new equipment production dominate production needs, compared with replacements. Conversely, the renewal steady state is characterized by $P_{tot}(t)$ approaching $K/\tau_{EoL}$, reflecting a state where production primarily focuses on replacing equipment reaching their EoL.

Numerical simulations employing a logistic curve for the in-use capacity, $Capacity(t)$, and a Weibull distribution for the EoL probability density function, $p_{EoL}(\theta)$, reveal the existence of two qualitatively distinct production behaviors. These behaviors depend on the interplay between the deployment characteristic time, $\tau_{dep}$, and the EoL average lifespan, $\tau_{EoL}$. 

\begin{figure*}[h!]
\begin{subfigure}{\linewidth}
\adjustbox{valign=c}{%
\begin{minipage}[t]{.05\linewidth}
\caption{}
\label{fig:Capacity}
\end{minipage}}%
\adjustbox{valign=c}{%
\begin{minipage}[t]{.95\linewidth}
  \centering
  \includegraphics[width = \textwidth]{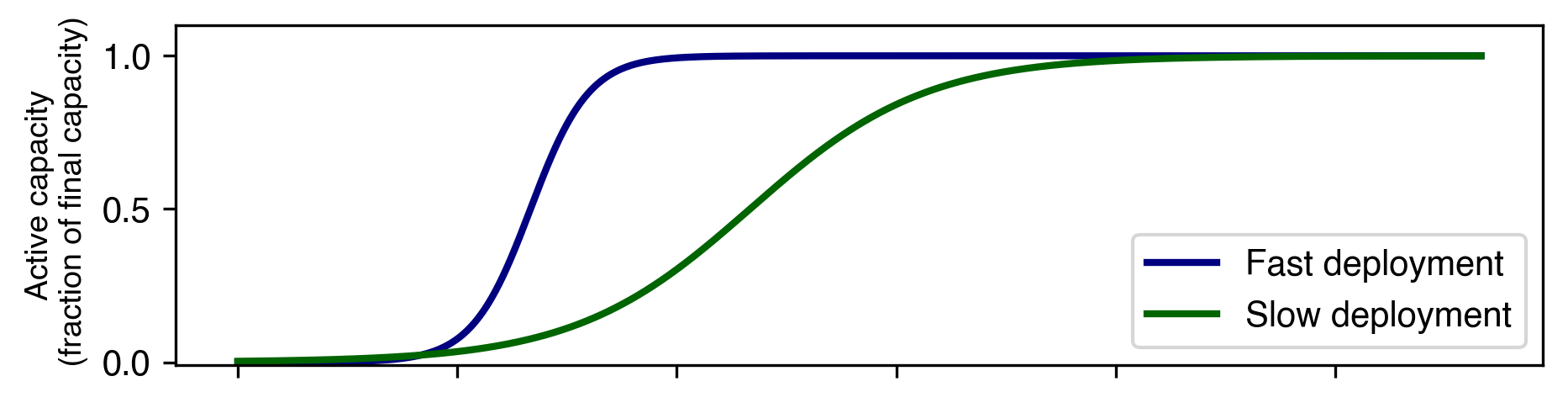}
\end{minipage}}
\end{subfigure}
\begin{subfigure}{\linewidth}
\adjustbox{valign=c}{%
\begin{minipage}[t]{.05\linewidth}
\caption{}
\label{fig:FastProduction}
\end{minipage}}%
\adjustbox{valign=c}{%
\begin{minipage}[t]{.95\linewidth}
  \centering
  \includegraphics[width = \textwidth]{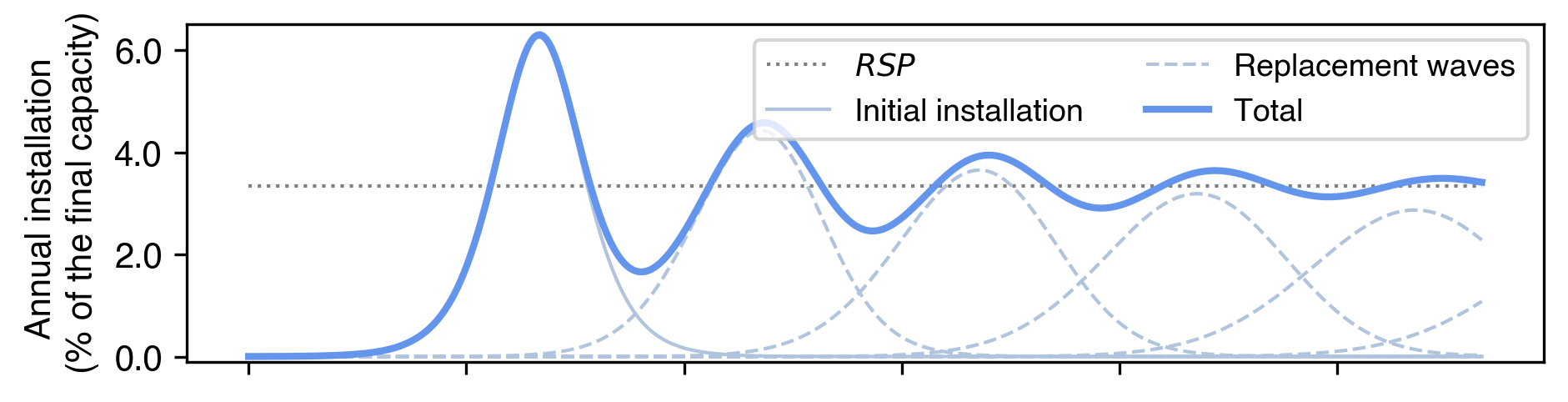}
\end{minipage}}
\end{subfigure}
\begin{subfigure}{\linewidth}
\adjustbox{valign=c}{%
\begin{minipage}[t]{.05\linewidth}
\caption{}
\label{fig:SlowProduction}
\end{minipage}}%
\adjustbox{valign=c}{%
\begin{minipage}[t]{.95\linewidth}
  \centering
  \includegraphics[width = \textwidth]{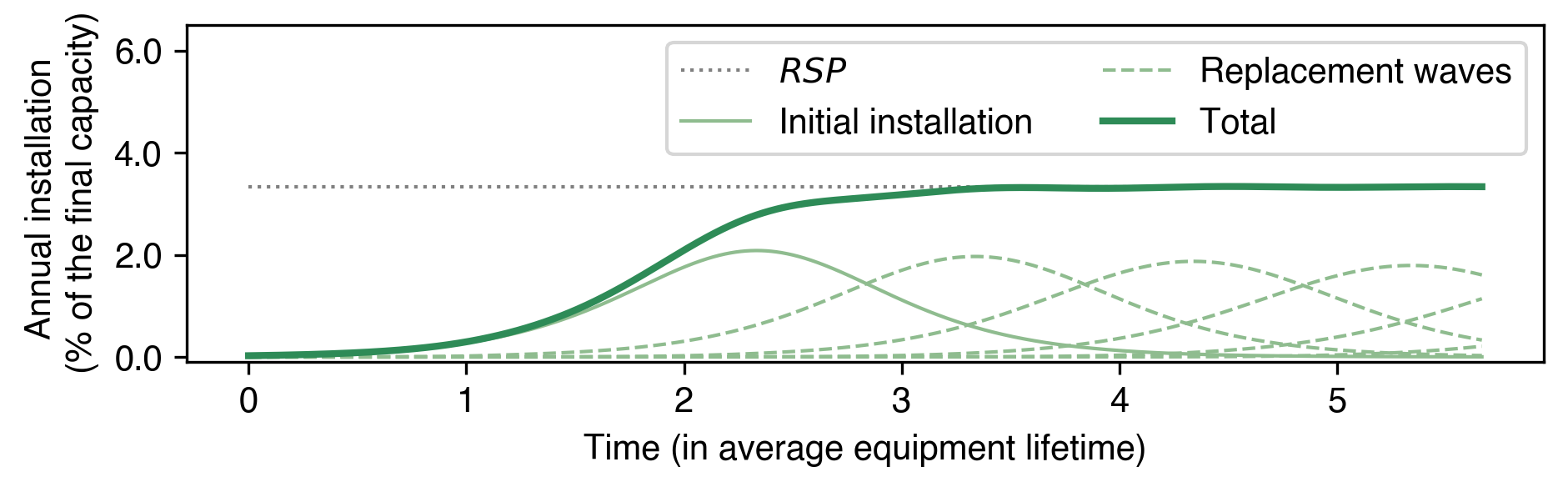}
\end{minipage}}
\end{subfigure}
\caption{\textit{Fast deployment} and \textit{Slow deployment} behaviors. Model outputs simulates a discretized version of Eq. (\ref{eq:model}) with a $0.1$ year step and tools from $numpy$ package within the Python programming environment. All the curves are computed with $\tau_{EoL} = 30$ years and $CV_{EoL} = 0.2$, for the \textit{fast deployment} $\tau_{dep} = 4$ years (\ref{fig:FastProduction}) and for \textit{slow deployment} $\tau_{dep} = 12$ years (\ref{fig:SlowProduction}).}
\label{fig:TwoBehaviors}
\end{figure*}

Figure~\ref{fig:TwoBehaviors} illustrates the two distinct production behaviors. The upper panel (\ref{fig:Capacity}) depicts the normalized in-use equipment (active capacity) for a \textbf{fast deployment} scenario (blue curve) and a \textbf{slow deployment} scenario (green curve). Panels (\ref{fig:FastProduction}) and (\ref{fig:SlowProduction}) show the total equipment production (bold line) for each deployment scenario, which is the sum of deployment production (lighter line) and the contributions from equipment replacement waves (dashed lines).

These behaviors can be summarized as follows:

\begin{itemize}
    \item \textbf{Fast Deployment}, see Fig. \ref{fig:Capacity} (blue curve) and Fig. \ref{fig:FastProduction}:
    \begin{enumerate}
        \item The system reaches its final capacity in a \textbf{shorter timeframe}.
     \item There is a \textbf{significant overshoot} in production compared to its steady renewal level, see Fig. \ref{fig:FastProduction}. 
     \item The transition from deployment to renewal steady state exhibits \textbf{damped oscillations} over a \textbf{long} period, see Fig. \ref{fig:FastProduction}.
    \end{enumerate}
\item \textbf{Slow Deployment}, see Fig. \ref{fig:Capacity} (green curve) and Fig. \ref{fig:SlowProduction}:
\begin{enumerate}
    \item The system reaches its final capacity over a \textbf{longer timeframe}, see Fig. \ref{fig:Capacity}.
    \item There is \textbf{no overshoot} in production compared to its steady renewal level level, see Fig. \ref{fig:SlowProduction}.
    \item The transition from deployment to renewal steady state is \textbf{monotonic and smooth}, see Fig. \ref{fig:SlowProduction}.
    \end{enumerate}
\end{itemize}

\subsubsection{Criterion for the existence of a production Overshoot}

The phenomenon of production overshoot is a hallmark of \textbf{fast deployment} scenarios. In simpler terms, when the deployment process is rapid, the production must surpass its steady renewal level to meet the swiftly growing active equipment capacity.

Formally, the \textbf{fast deployment} regime is characterized by:

$$max(P_{tot}) > P_{MSS} = \frac{K}{\tau_{EoL}}$$

While the precise value of $max(P_{tot})$ requires numerical computation, a simple lower bound can be established using $P_{dep}(t_{peak})$, the peak deployment production. Leveraging the logistic growth equation (Eq. (\ref{eq:proddep})), this lower bound translates to:

\begin{equation} \label{eq:order0}
    \frac{K}{4\tau_{dep}}>\frac{K}{\tau_{EoL}} \implies \frac{\tau_{dep}}{\tau_{EoL}} < \frac{1}{4}
\end{equation}

This equation implies that an overshoot will occur when the characteristic deployment time ($\tau_{dep}$ ) is less than one-quarter of the average equipment lifespan ($\tau_{EoL}$).

This lower bound can be refined by adding the contribution of the first replacement wave:
\begin{equation} \label{eq:overshoot}
\begin{split}
        max(P_{tot}(t))&\ge P_{dep}(t_{peak}) + R_{1}(t_{peak})\\
        &\ge \frac{K}{\tau_{dep}} \left(\frac{1}{4} + \int_{0}^{+\infty}\frac{e^{\frac{\theta}{\tau_{dep}}}}{\left(1+e^{\frac{\theta}{\tau_{dep}}}\right)^2}p_{EoL}(\theta)d\theta \right)
\end{split}
\end{equation}

\begin{figure*}[h]
    \begin{subfigure}{.5\textwidth}
    \centering
        \includegraphics[width=\linewidth]{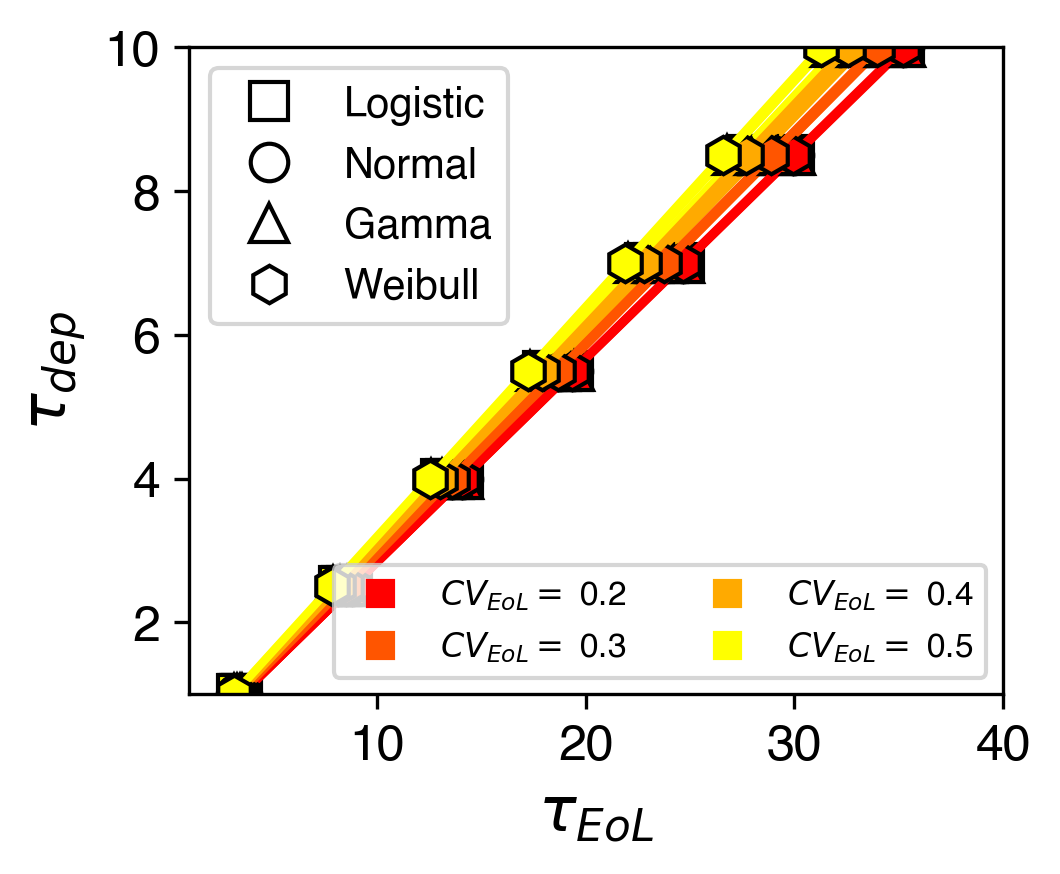}
        \captionsetup{width=0.9\linewidth}
        \caption{Couple ($\tau_{EoL}$, $\tau_{dep}$) beyond which there is an overshoot, for various distribution $p_{EoL}$ and various values of $CV_{EoL}$. The space under the curve represents the parameter values for which there is an overshoot.}
        \label{fig:Overshoot}
    \end{subfigure}%
    \begin{subfigure}{.5\textwidth}
    \centering
        \includegraphics[width=\linewidth]{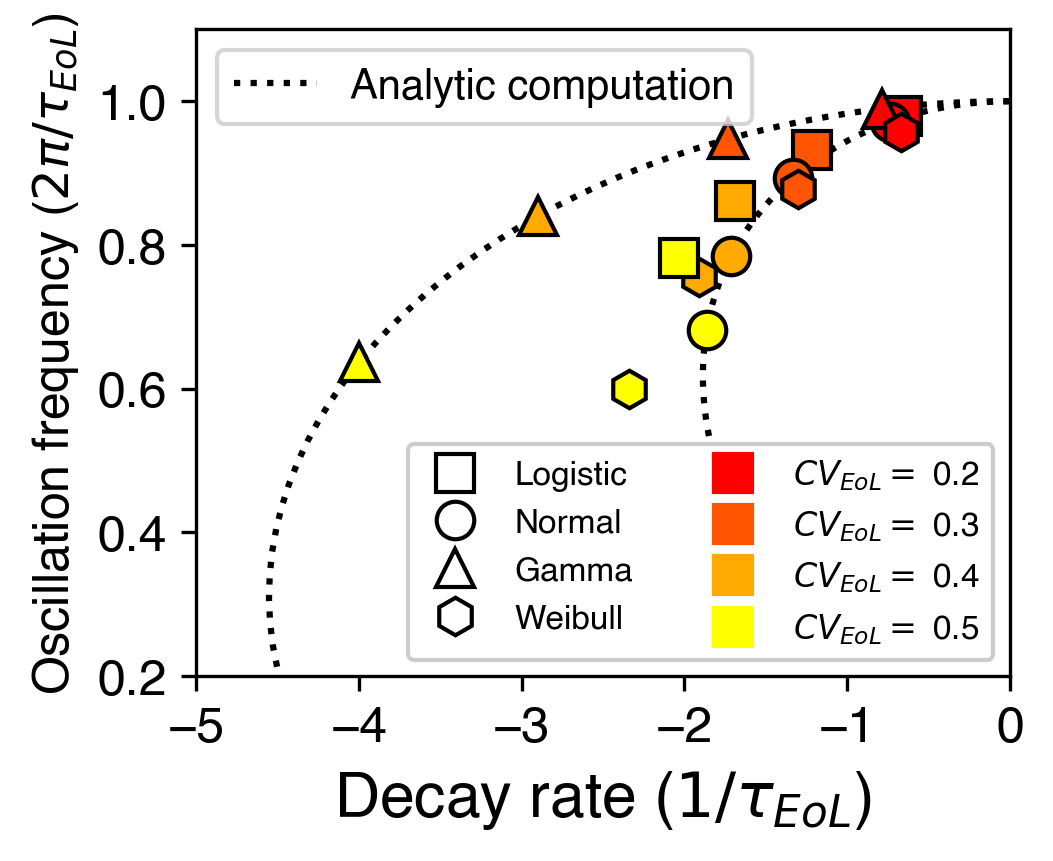}
        \captionsetup{width=0.9\linewidth}
        \caption{Location of the non-zero root of Eq. (\ref{eq:carac}) with higher real part, for different distributions and $CV_{EoL}$ values. These roots captures the oscillatory transient behavior : the real part corresponds to the decay rate and the imaginary part to the frequency.}
        \label{fig:Roots}
    \end{subfigure}
    \caption{Overshoot and roots of the characteristic equation for different distributions and values of $CV_{EoL}$}
\end{figure*}

Figure~\ref{fig:Overshoot} depicts the regions in the parameter space, defined by the deployment characteristic time ($\tau_{dep}$ ) and EoL average lifespan ($\tau_{EoL}$ ), where the condition $P_{dep}(t_{peak}) + R_{1}(t_{peak}) = RSP$ is satisfied. As the data points align along straight lines, this suggests that for a given coefficient of variation of EoL ($CV_{EoL}$ ) and EoL probability density function ($p_{EoL}$ ), the \textbf{fast/slow deployment behavior} is solely determined by the ratio $\tau_{dep}/\tau_{EoL}$. 

It is important to note that Figure~\ref{fig:Overshoot} relies on the approximation $P_{dep}(t_{peak}) + R_{1}(t_{peak})$, not the exact maximum production, $max(P_{tot})$. However, as demonstrated in Figure~\ref{fig:TwoBehaviors}, this approximation proves to be highly accurate even for slow deployment scenarios where the contribution of the second replacement wave is negligible.

Based on this analysis, we can conclude that for a wide range of $CV_{EoL}$ values and any $p_{EoL}$ function, the critical ratio, denoted by $\mathbf{r_c} = \tau_{dep}/\tau_{EoL}$, that separates \textbf{fast behavior} from \textbf{slow behavior} lies within the range [0.27, 0.34]. As $\tau_{dep}$ is implicitly defined through the canonical form of the logistic function, these $r_c$ values can be hard to read. In a more practical form, this corresponds to a maximum active equipment capacity growth over one average equipment lifespan of around $70\%$ of the final capacity.

Furthermore, Equation (\ref{eq:overshoot}) suggests that the \textbf{intensity of the production overshoot} is directly proportional to the ratio $\tau_{EoL}/\tau_{dep}$ with additional, gradually diminishing corrective terms accounting for contributions from subsequent replacement waves.

\subsubsection{Characterization of the Transient Dynamics towards Renewal Steady State}

The second critical industrial concern is the \textbf{duration of the transient phase} before reaching the Renewal Steady State (RSS). In essence, the question is: \textbf{For how long will production oscillate after peak deployment is achieved?}

As observed in Figure~\ref{fig:TwoBehaviors}, the production oscillations exhibit a \textbf{damped} behavior, with annual production converging exponentially towards the renewal steady production, $RSP$. This convergence phenomenon can be rigorously proven by applying specific techniques from renewal theory, as outlined in classical works by Leadbetter (1964) \cite{leadbetter1964} and Cox (1970) \cite{cox1970}. In our specific case, these techniques will be employed to construct an explicit solution to Eq. (\ref{eq:model}), thereby formally demonstrating the exponential convergence.

Applying a Laplace transformation to Eq.(\ref{eq:model}) and noting the Laplace transforms of $f(t) \leftrightarrow f^{\ast}(p)$, we get :

\begin{equation} \label{eq:laplaceModel}
    P_{tot}^{\ast}(p) = P_{dep}^{\ast}(p) + P_{tot}^{\ast}(p)\times p_{EoL}^{\ast}(p)
\end{equation}

The Laplace transform of $P_{tot}(t)$ is then :

\begin{equation}\label{eq:laplacePtot}
    P_{tot}^{\ast}(p) = \frac{P_{dep}^{\ast}(p)}{1 - p_{EoL}^{\ast}(p)}
\end{equation}

The exponential rate of convergence is given by the roots of the \textit{characteristic equation} :

\begin{equation} \label{eq:carac}
    p_{EoL}^{\ast}(p) = 1
\end{equation}

Using the following inversion formula \cite{widder2015laplace}, with $c$ such as all singularities lie to the left of the line of integration :
\begin{equation} \label{eq:inversion}
    f(t) = \frac{1}{2\pi i}\int_{c - i\infty}^{c+i\infty} e^{pt}f^{\ast}(p)dp
\end{equation}

Assuming that function (\ref{eq:laplacePtot}) is holomorphic, applying Cauchy's residue theorem and the inversion (\ref{eq:inversion}) to Eq. (\ref{eq:laplacePtot}) gives :
\begin{equation*} 
    P_{tot}(t) = \sum \left\{\text{Residues of }
    \frac{P_{dep}^{\ast}(p)}{1 - p_{EoL}^{\ast}(p)}e^{pt}\right\}
\end{equation*}

Then, assuming that $p_{i}$ - the roots of Eq. (\ref{eq:carac}) - are simple, which from a practical point of view will very often be verified or it will be easy to slightly modify $p_{EoL}$ so that they are :

\begin{equation*} 
    P_{tot}(t) = \sum_{p_{i}} e^{p_{i}t}\times \frac{P_{dep}^{\ast}(p_{i})}{-p_{EoL}^{\ast'}(p_{i})}
\end{equation*}

Finally, since $p_{EoL}$ is a probability density then $0$ is a root of Eq. (\ref{eq:carac}):

\begin{equation*}
\left\{
    \begin{array}{lll}
        p_{EoL}^{\ast}(0) = \int_{0}^{+\infty} p_{EoL}(t)dt = 1\\[5pt]
        p_{EoL}^{\ast'}(0) = \int_{0}^{+\infty} tp_{EoL}(t)dt = \tau_{EoL}\\[5pt]
        P_{dep}^{\ast}(0) = \int_{0}^{+\infty} P_{dep}(t)dt = K\\
    \end{array}
\right.
\end{equation*}

Separating the $0$ pole, corresponding to $RSP$, the final form is :

\begin{equation}\label{eq:Ptotfinalform}
    P_{tot}(t) = \frac{K}{\tau_{EoL}} + \sum_{p_{i} \neq 0} e^{p_{i}t}\times \frac{P_{dep}^{\ast}(p_{i})}{-p_{EoL}^{\ast'}(p_{i})}
\end{equation}

Noting $u_{max} \pm i \, v_{max}$ the couple of roots with the higher real part $u_{max}$ (displayed in Figure (\ref{fig:Roots})), we approximate $P_{tot}(t)$ by the value of the less damped residue contribution:

\begin{equation}\label{eq:Ptotapprox}
    P_{tot}(t) = \frac{K}{\tau_{EoL}} + e^{u_{max}t}\left\vert \frac{P_{dep}^{\ast}(u_{max} + iv_{max})}{-p_{EoL}^{\ast'}(u_{max} + iv_{max})}\right\vert cos(v_{max}t + \phi)
\end{equation}

\begin{figure*}[h]
    \begin{subfigure}{.5\textwidth}
    \centering
        \includegraphics[width=\linewidth]{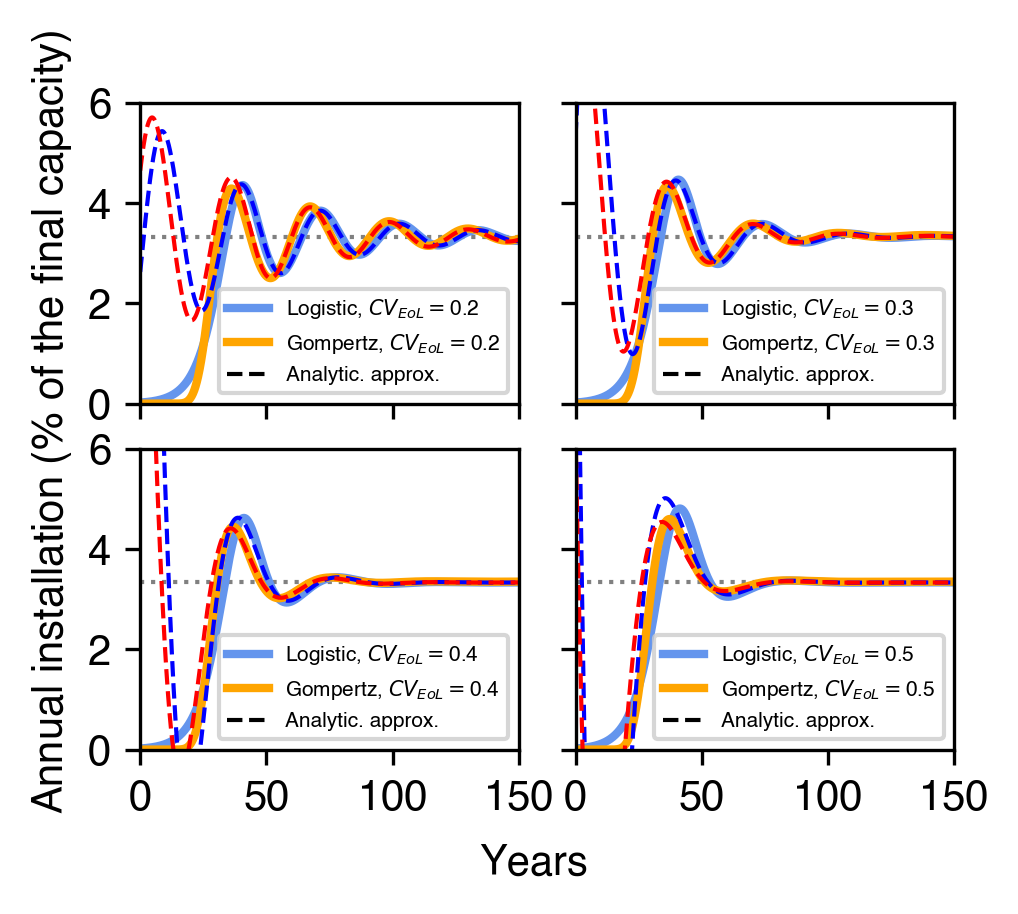}
        \captionsetup{width=0.9\linewidth}
        \caption{Comparison of the analytical first residue approximation given by (Eq. \ref{eq:Ptotapprox}) with the simulated output for the fixed values $\tau_{dep}=5$ y.,$\tau_{eol}=30$ y. and different values of $CV_{EoL}$. }
        \label{fig:1stResidueApp}
    \end{subfigure}%
    \begin{subfigure}{.5\textwidth}
    \centering
        \includegraphics[width=\linewidth]{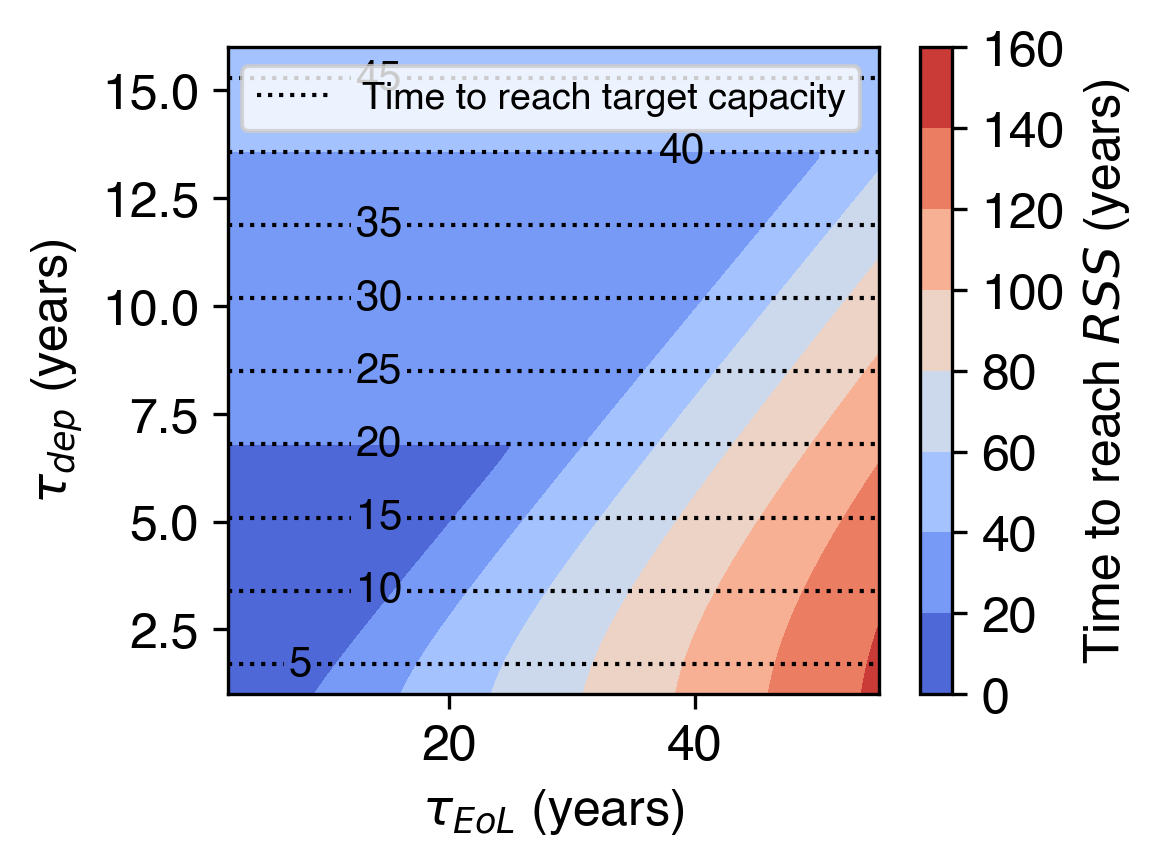}
        \captionsetup{width=0.9\linewidth}
        \caption{Transition time to reach \textit{RSS} (Renewal Steady State) as a function of $\tau_{dep}$ and $\tau_{EoL}$. The duration is computed from (Eq. \ref{eq:Ptotapprox}) using $CV_{EoL}=0.3$ and \textit{RSS} considered reached when $|P_{tot}(t) - RSP|<0.05$ and $|Capacity(t) - K|<0.05$ }
    \label{fig:TransitionTime}
    \end{subfigure}%
    \caption{Effectiveness of the analytical approximation (Eq. \ref{eq:Ptotapprox}) and corresponding transition time.}
\end{figure*}

Our analysis reveals two key effects influencing the convergence towards the Renewal Steady State (RSS):

\begin{itemize}
    \item Impact of coefficient of variation of EoL (CV$_{EoL}$): A lower CV$_{EoL}$ implies a narrower distribution of equipment EoL lifespan. In practical terms, this signifies that most equipment units are likely to reach EoL close to the average lifespan, $\tau_{EoL}$. Consequently, the production peak will be replicated, albeit with damped intensity, at intervals of approximately $\tau_{EoL}$ years due to concentrated replacement waves. Conversely, a higher CV$_{EoL}$ broadens the EoL distribution, leading to \textbf{more dispersed replacement peaks} and a faster convergence towards the steady-state production level (higher damping rate).

    \item Approximation of damping rate using Normal distribution: The real parts of the roots of Eq. (\ref{eq:carac}) for various EoL probability density functions (pdfs) appear to be very similar. This implies that the damping rate calculated analytically for a normal distribution serves as a good approximation regardless of the actual EoL pdf. The accuracy of this approximation is particularly pronounced for low CV$_{EoL}$ values. However, for higher CV$_{EoL}$ values, the Gamma distribution might exhibit a slightly higher damping rate. While the precise reason for this discrepancy is not entirely clear, it could potentially be explained by a more nuanced analysis considering the skewness of the distribution. Nonetheless, the objective here is to explain the dynamics using the simplest and most efficient parameterization.
    
\end{itemize}

Figure~\ref{fig:1stResidueApp} demonstrates that the \textbf{first residue approximation} (Eq.\ref{eq:Ptotapprox}) rapidly gains accuracy. This occurs because subsequent terms in the solution exhibit faster damping due to their roots having lower real parts. This observation allows us to develop an \textbf{approximate estimate for the transition duration}.

Figure~\ref{fig:TransitionTime} depicts the duration of the transient phase between the peak deployment level and the Renewal Steady State (RSS), characterized by a constant capacity of $K$ and annual production of $RSP$. In Figure~\ref{fig:TransitionTime}, the transition duration is precisely defined as the time needed for both $|P_{tot}(t) - RSP|$ and $|Capacity(t) - K|$ to fall below $5\%$ of their respective steady-state values ($RSP$ and $K$). It is important to note that this $5\%$ threshold is an adjustable parameter that influences the measured duration; a higher threshold will yield a shorter apparent transition time.

Figure~\ref{fig:TransitionTime} underscores the existence of a \textbf{trade-off} for a given EoL distribution (characterized by the pair $(\tau_{EoL}, CV_{EoL})$) between the time required to reach the target capacity (dotted line) and the overall transition duration. Achieving the target capacity swiftly necessitates fast deployment, which leads to a significant deployment peak, inducing production oscillations and consequently, a longer transition period. 

\subsection{The fast deployment dilemma}

The previous model, despite its simple parameterization, captures an essential issue of a \textit{deployment to renewal} dynamic regarding the annual equipment unit production : the existence of a trade-off between the time to reach the target capacity and the time to reach a steady renewal production.

On the one hand :
\begin{itemize}
    \item The time to reach target capacity is \textbf{proportional} to the characteristic time of deployment time $\tau_{dep}$ (as this is what measures this parameter). 
    \item The deployment production peak intensity is \textbf{inversely proportional} to $\tau_{dep}$.
\end{itemize}

On the other hand :
\begin{itemize}
    \item The annual production in a renewal steady state ($RSP$) is inversely proportional to the average lifespan of an equipment $\tau_{EoL}$.
\end{itemize}

So there is a point at which reaching target capacity more quickly means that deployment peak production exceeds renewal steady production and thus creates oscillation in the production. This is the \textit{fast deployment dilemma}, after this critical point ($\tau_{dep}=r_{c}\tau_{EoL}$), a gain in time on reaching the target capacity costs a longer transition and more intense oscillations, and vice versa.

The benefits of reaching target capacity quickly are often obvious as the technology answer a particular need or demand. In the other hand the cost of this fast deployment, understood as a annual production undergoing overshoot and oscillations, can be underestimated. Indeed as it will be shown in the following sections the overshoot and the oscillations can be of huge magnitude, and such an equipment production dynamic can lead to different risks.

\subsection{The role of lifespan variance}

Previous findings must be nuanced by the significant role of equipment lifespan variance, captured in the model by the parameter $CV_{EoL}$. Indeed Figure \ref{fig:TransitionTime} shows that if the deployment peak remains unchanged, oscillations can be significantly damped for some $CV_{EoL}$ values.

The dynamic is the following : an equipment lifespan precisely determined will result in concentrated replacement waves, echoes of the production peak, and thus a long and fluctuating transition. On the contrary, a large $CV_{EoL}$ will spread out the replacement waves and the transition will be short and smooth. 

Increasing $CV_{EoL}$ thus seems a way out the fast deployment dilemma, reducing the cost to only an important overshoot. However tuning $CV_{EoL}$ would not be that easy in practice as it must be done at a constant $\tau_{EoL}$ : for every equipment decommissioned in advance, an other equipment decommission must be delayed. We can easily imagine the problems and difficulties of implementing such a strategy. 

\section{Case studies: Applying the modeling framework to Nuclear power plants and Smartphones}
\label{sec:CaseStudies}
This section outlines the methodology employed to compare the dynamics predicted by the model, based on parameters characterizing the deployment S-curve and the lifespan distribution, with the historical production data from technological case studies displayed in Section \ref{sec:ComparisonIntro}. For each, nuclear power plants and smartphones, the following steps were undertaken:

\begin{itemize}
\item $Capacity(t)$ and $P_{tot}(t)$ definitions: Identify appropriate metrics to quantify equipment capacity and annual production. Capacity should reflect the number of equipment units in service over time, while production should represent the annual output of equipment units.

\item Deployment curve fitting: Use historical data on capacity to fit a logistic deployment curve. This process will determine model parameters such as the final capacity ($K$), deployment characteristic time ($\tau_{dep}$), and peak deployment time ($t_{peak}$). 

\item EoL distribution determination: Identify or estimate a range of values for the EoL distribution parameters ($\tau_{EoL}$, $CV_{EoL}$) based on existing literature or available data.

\item Model comparison: Compare the annual production curve generated by the model with historical production data. This comparison serves to assess the model ability to explain real-world production trends.
\end{itemize}

\subsection{Nuclear power plants - a fast deployment example}

The global deployment of nuclear power plants exemplifies the \textit{deployment to renewal} dynamics explored in this analysis. Since the 1990s, global nuclear capacity, measured in Gigawatt (GW) to reflect total power generation capability, has exhibited minimal growth~\cite{nuclearsource}. This indicator corresponds directly to the technological service, energy production, which is not the case for the number of active power plants. Indeed, Nuclear power plants are still undergoing a phase of continuous technological improvements and the nominal power output of a single reactor has more than double between the 1970s and the 2010s.

For this case study, \textbf{production} data will represent the annual capacity newly connected to the electric grid to be coherent with the capacity measurement. Determining a definitive average lifespan for nuclear facilities is challenging, and the literature lacks a universally accepted estimate. Based on available data, we adopt a conservative value of 50 years for the equipment lifespan (\textbf{$\tau_{EoL}$}) in this analysis.

\begin{table}[h]
\centering
\begin{tabular}{|c|c|c|c|c|}
    \hline
    & \multicolumn{4}{|c|}{Capacity deployment (fit)} \\
    \hline
    Parameters & $K$ (GW) & $\tau_{dep}$ (y.) & $t_{peak}$ & $R^{2}$\\
    \hline
    Value & $363.5$ & $4.35$ & $1982$ & $0.998$ \\
    \hline
\end{tabular}

\vspace*{0.5 cm}

\centering
\begin{tabular}{|c|c|c|}
    \hline
    &  \multicolumn{2}{|c|}{End-of-Life (estimated)}\\
    \hline
    Parameters & $\tau_{EoL}$ (y.) & $CV_{EoL}$ \\
    \hline
    Value & $50$ [$45-55$] & $0.2$ [$0.15-0.5$]\\
    \hline
\end{tabular}
\caption{This table summarizes the parameters used in the nuclear deployment model. The capacity deployment parameters are fitted using optimization techniques available in the $scipy$ package within the Python programming environment. The parameters regarding EoL are estimated from available data. The range of parameters between brackets is used to measure model sensitivity.}
\label{tab:nuclear}
\end{table}

\begin{figure*}[!h]
\centering
\includegraphics[width=\linewidth]{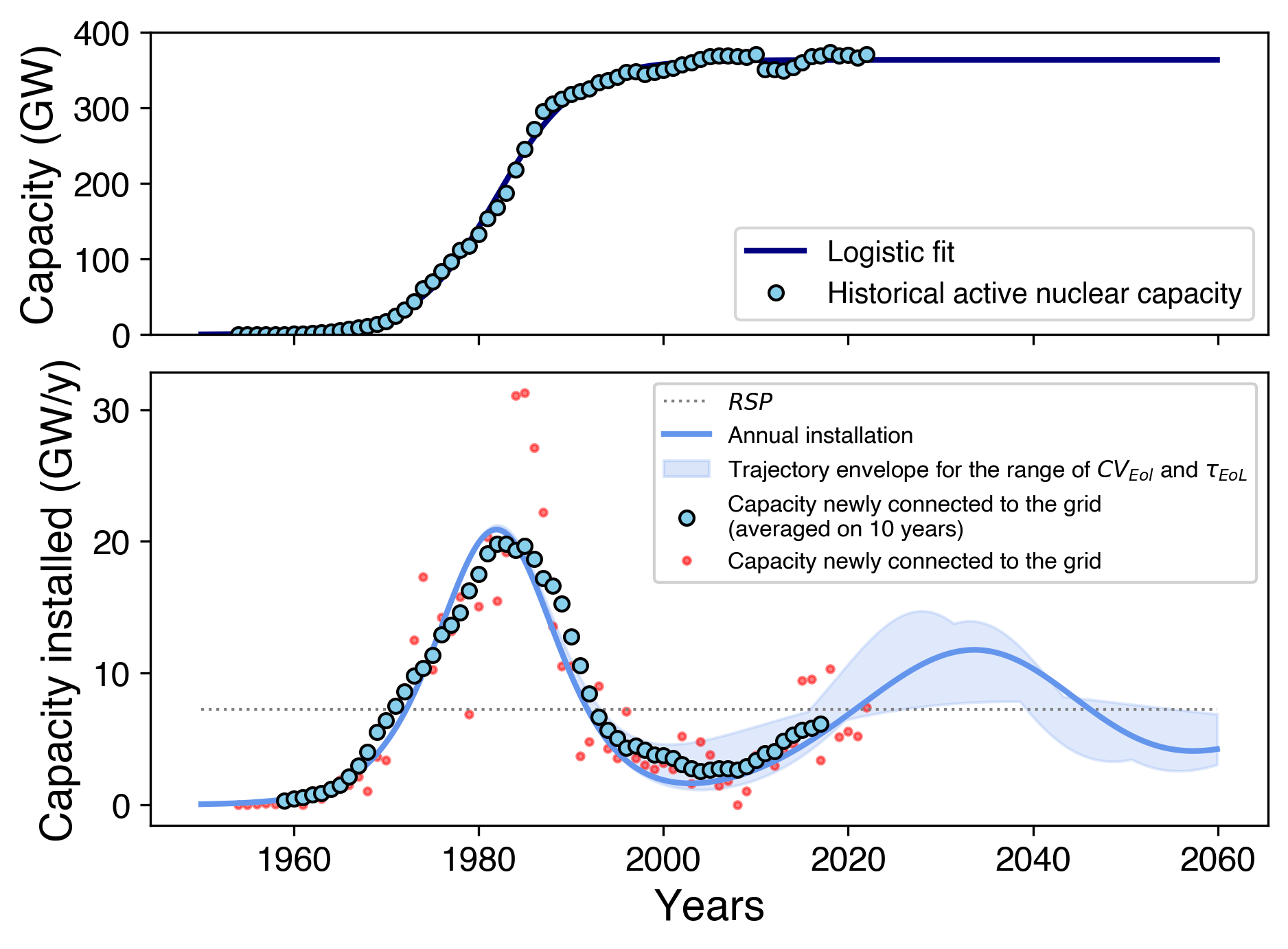}
\caption{The figure shows a comparison between historical data on global nuclear capacity and annual new plant connections (production) and the model output for production. The model output is generated using parameters from Table~\ref{tab:nuclear}.}
\label{fig:Nuclear}
\end{figure*}

The annual production curve generated by the model (light blue line, lower panel of Figure~\ref{fig:Nuclear}) exhibits overshoot and subsequent damped oscillations. This behavior is indicative of a fast deployment scenario, as the characteristic deployment time ($\tau_{dep}$) of 4.35 years is significantly lower than the critical value of $r_c \times \tau_{EoL}$ (approximately 13.5 years, with $r_c = 0.27$ and $\tau_{EoL} = 50$ years).

The historical production data (blue dots, lower panel of Figure~\ref{fig:Nuclear}) demonstrates good agreement with the model predicted production trajectory based on the historical capacity data. A significant overshoot is observed, with the deployment peak exceeding steady renewal production by nearly a factor of three. Additionally, production exhibits a decline of almost 90\% between 1982 and the 2000s. These findings indicate that this relatively simple model effectively captures some key drivers of the \textit{deployment to renewal} dynamics in nuclear power plant deployment.

\subsubsection*{The case of nuclear power plant : specific insights and limitations}

The application of the model to nuclear power plant deployment presents three key considerations compared to a generic technological deployment scenario.

\textbf{Limited population and averaging:} due to the relatively small number of active nuclear power plants (fewer than 500 globally), the modeled production curve reflects the expected value of added capacity rather than the exact annual GW increase. Consequently, the model predictions are compared to 10-year averaged historical data (blue dots) instead of raw annual data (red dots in Figure~\ref{fig:Nuclear}). This is further necessitated by the variability in power output capacity between individual plants, which can range from single to double units.

\textbf{Plant refurbishment and lifespan extension:} The model incorporates the concept of an EoL distribution but does not explicitly account for the possibility of extending operational lifespan through plant refurbishment. While not explicitly modeled, the chosen average lifespan of 50 years appears to yield reasonable results when compared to historical data.

\textbf{Nuclear skills gap and fast deployment risks:} An additional point of interest is the current state of the nuclear power industry, which is experiencing its first period of declining activity. Concerns regarding a potential nuclear skills gap have been raised in the literature~\cite{international2004nuclear,cc2020EPR}. This situation highlights the potential risks associated with fast deployment strategies in the nuclear sector, as a skilled workforce is crucial for plant operation and maintenance.

\subsection{Smartphones - a slow deployment example}

This subsection presents the second case study exhibiting a qualitatively different production behavior : global smartphone deployment. Here, annual \textbf{production} data is estimated based on the number of smartphones sold worldwide each year, acknowledging potential limitations in data source credibility~\cite{smartphonesource,2handsales2014,2handsales2015/16,2handsales2017,2handsales2018/19,2handsales2020,2handsales2021,2handsales2022}. Determining the average lifespan of a smartphone is also challenging. However, based on available production data, a value of 3.7 years appears to yield the most consistent results. It is important to recognize that the ``true" average lifespan may reside within a range around 3 years. This characteristic positions smartphone deployment as a \textbf{slow deployment} scenario within the framework of the model.

\begin{table}[h]
\centering
\begin{tabular}{|c|c|c|c|c|}
    \hline
    & \multicolumn{4}{|c|}{Capacity deployment (fit)} \\
    \hline
    Parameters & $K$ (Billions) & $\tau_{dep}$ (y.) & $t_{peak}$ & $R^{2}$\\
    \hline
     Value & $6.94$ & $2.49$ & $2015$ & $0.995$ \\
    \hline
\end{tabular}

\vspace*{0.5 cm}

\centering
\begin{tabular}{|c|c|c|}
    \hline
    &  \multicolumn{2}{|c|}{End-of-Life (estimated)}\\
    \hline
    Parameters & $\tau_{EoL}$ (y.) & $CV_{EoL}$ \\
    \hline
    Value & $3.7$ [$3-6$] & $0.3$ [$0.2-0.5$]\\
    \hline
\end{tabular}
\caption{This table summarizes the parameters used in the smartphone deployment model. The capacity deployment parameters are fitted using optimization techniques available in the $scipy$ package within the Python programming environment. The parameters regarding EoL are estimated from available data. The range of parameters between brackets is used to measure model sensitivity.}
\label{tab:smartp}
\end{table}


\begin{figure*}[!h]
\centering
\includegraphics[width=\linewidth]{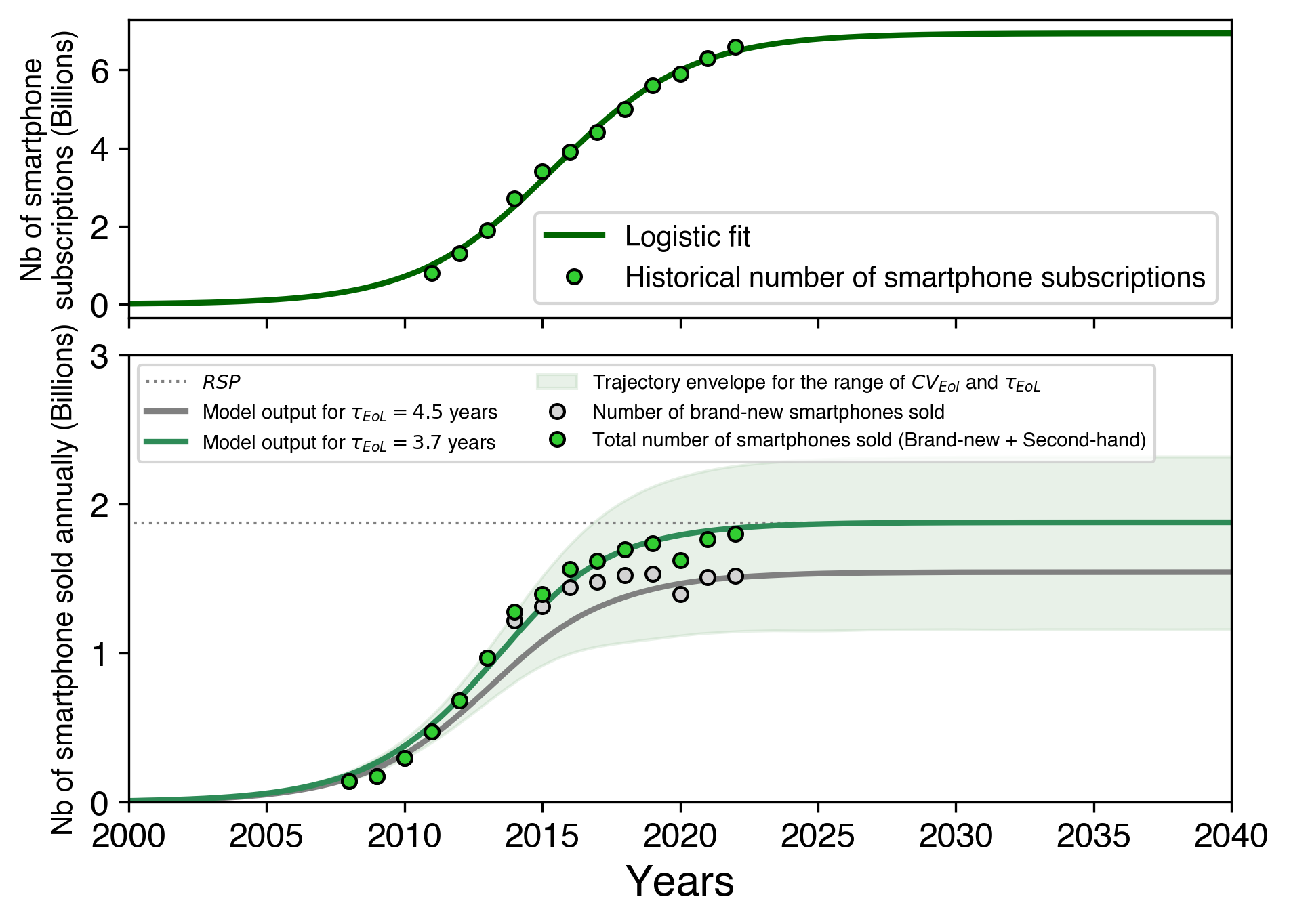}
\caption{The figure shows a comparison between historical data on global smartphone subscriptions (capacity) and annual smartphone sales (production) alongside the model output for production generated using parameters from Table~\ref{tab:smartp}. More precisely, the grey curve corresponds an output with a renewal production consistent with new sales values (grey dots) and the green curve corresponds to an output consistent with total sales values, brand-new and second-hand (green dots).}
\label{fig:Smartphones}
\end{figure*}

The analysis reveals a \textbf{slow deployment} behavior for smartphones despite a potentially shorter deployment characteristic time ($\tau_{dep}$) compared to the nuclear power plant case. This seemingly counterintuitive observation can be attributed to the critical ratio $\tau_{dep}/\tau_{EoL}$. In the context of smartphones, the average equipment lifespan ($\tau_{EoL}$) is significantly shorter than that of nuclear power plants. Consequently, even with a potentially faster deployment process (lower $\tau_{dep}$), the critical ratio remains high, leading to slow deployment dynamics according to the model predictions.

The model agreement with historical data should not be seen as an uncompromising validation of its predictive capabilities. Unreliable access to production data and the arbitrary choice of an average lifespan make the quantitative aspect of this comparison more fragile. The strength of the model lies in its ability to explain the qualitative difference between the two production profiles with coherent parameters, in particular the lifetime of around 3 years. In the case of smartphones, the average lifespan is short enough for production not to fluctuate.

\subsubsection*{The case of smartphones : specific insights and limitations}

Unlike nuclear power plants, the significantly larger population of smartphones allows for the confident application of the law of large numbers. However, a discrepancy is observed between the historical data on \textbf{item production} (number of new smartphones sold annually, represented by gray dots in Figure~\ref{fig:Smartphones}) and the model output. While a production plateau observed in the 2020s suggests an average equipment lifespan ($\tau_{EoL}$) of 4.5 years, this value results in a poor fit for the historical data in the 2010s. Conversely, a $\tau_{EoL}$ of slightly less than 4 years yields a good fit for the 2010s data, but the model overestimates the steady-state production level ($P_{MSS}$).

This discrepancy can be addressed by incorporating the effect of \textbf{second-hand sales} (represented by green dots in Figure~\ref{fig:Smartphones}), which became increasingly significant in the late 2010s. Two potential modelizations of this effect are presented:

\textbf{Static model:} We assume a constant equipment lifespan ($\tau_{EoL}$). When a smartphone is sold second-hand, it is treated as new production within the model. This approach results in the green curve, which exhibits a good fit with the data.

\textbf{Dynamic model:} We consider the rise of second-hand sales to effectively extend the average equipment lifespan ($\tau_{EoL}$) over time. This dynamic effect would necessitate a shift from the green curve to the gray curve during the 2010s to reflect the increasing contribution of second-hand devices.

The impact of second-hand markets highlights a potential future extension of the model. Incorporating a time-dependent EoL distribution, where the average $\tau_{EoL}$ can increase over time, could offer a more nuanced representation of deployment dynamics in markets with significant second-hand activity.

\section{Discussion}
\label{sec:Discussion}

In addition to illustrating the two types of behavior and the constraints associated with ``fast deployment", these two case studies offer an interesting reflection on the factors that effectively constrain dynamics in practical applications. Within our theoretical framework, production dynamics are dictated by two given factors that appear as constraints: the deployment curve and the lifetime. In practice, these two factors may be the result of external causes that are not independent of production dynamics.

The final capacity plateau may be a target value, set in advance by a player, or a saturation value, set by essential constraints. A country final nuclear power generation capacity depends on a political decision rather than an essential deployment constraint. On the contrary, the number of smartphones in service is limited by the country's total population.
Moreover, the lifespan of smartphones is defined less by the end of their useful life than by the purchase of a new model. This is not the case for nuclear power plants, whose lifespan is extended as far as possible.

This analysis suggests that for the deployment of long-life equipment such as public infrastructure, including nuclear power plants, providing a service is the main objective. Achieving this objective rapidly under equipment lifespan strict constraint comes at the cost of oscillating production, which translates into industrial cycles with all its associated obstacles. Regarding consumer goods, produced by companies pursuing economic viability, the situation is quite different. In this case, companies benefit from having the most stable production, and therefore ``slow deployment'' is the preferential option. To achieve this, manufacturers produce new versions of their equipment, as is the case with smartphones, thus reducing the ``effective'' lifespan of each piece of equipment. Lifespan can appear here as a kind of adjustment variable, enabling a ``slow deployment'' regime. However, the speed of incremental technological advances must also be taken into account, as this also accelerates renewal.

This discussion highlights two limitations and, simultaneously, two potential extensions of this ``static" model: a further deployment of capacity and a variable equipment lifespan. A capacity that stabilizes before undergoing a second wave of deployment would provide a more detailed understanding of the interactions between deployment and renewal, as well as investigating the relationship between plateau duration and oscillation intensity. An equipment lifespan that varies over time would help to better capture deployment dynamics such as those of smartphones, and offer industrial strategies additional room for maneuver, particularly with regard to mitigating oscillations.

Beyond these dynamic aspects, the model current framework focuses on a single technology. While the results produced are already interesting, practical applications where different technologies compete/cooperate in filling the same service could escape this single technology scope. Nuclear power generation, for example, is part of a national energy landscape which is characterized by a diverse mix of energy sources. Future extensions of this model could incorporate the influence of heterogeneous energy mixes on the deployment dynamics of individual technologies. This would involve taking explicit account of variations in total desired capacity, potentially induced by factors such as national energy policy, renewable energy penetration and overall energy demand. Incorporating these factors would enable the model to capture the interaction between different energy sources within a national grid, and provide a more nuanced understanding of deployment trajectories for specific technologies.

The oscillating nature of production in some scenarios raises industrial issues, highlighted in section \ref{sec:KeyTakeaways}, including the question of raw materials. These fluctuations in demand for raw materials could potentially disrupt supply chains or necessitate the building up of strategic stocks of critical materials. However, as technology undergoes a phase of incremental technological evolution, its material footprint could decrease with each renewal. Further research is needed to understand these dynamics and their long-term implications for material requirements.

\section{Implications for industrial deployment strategies}
\label{sec:IndustrialConsequences}

Our study demonstrates that oscillations in economic activity, such as business cycles, can arise endogenously from the dynamics of capital installation and renewal within specific industries. This finding contrasts with traditional explanations of cyclical industries, which often attribute fluctuations to exogenous factors in the broader macroeconomic environment.

Specifically, we have identified that the timing of major capital investments and subsequent renewal or replacement cycles can create periodic fluctuations in industry-specific economic activity. These endogenous cycles are driven by the internal dynamics of the industry itself, rather than being primarily responsive to external economic forces.

It is important to note, however, that the existence of these industry-specific, endogenous cycles does not preclude the influence of macroeconomic factors. In reality, both endogenous and exogenous cycles likely coexist and interact. The endogenous cycles we have identified may be modulated, amplified, or attenuated by broader economic trends, resulting in a complex interplay between industry-specific dynamics and macroeconomic conditions.

\subsection*{Key modeling features}

The modeling assumptions and parameters are summed up in Table~\ref{tab:param}. Two key assumptions underpin the model:
\begin{enumerate}
    
\item \textbf{Sustained technological capacity:} Following deployment, a certain level of technological capacity is maintained. This implies a period of incremental evolution with no disruptive technological breakthrough rendering the deployed technology obsolete. Mathematically, the evolution of active capacity is represented by an S-shaped curve.

\item  \textbf{Equipment lifespan distribution:} The active lifespan of individual equipment before reaching EoL follows a bell-shaped probability distribution centered around an average value. This distribution reflects inherent variations in equipment lifespans.
\end{enumerate}

The model uses a logistic curve to capture the S-shaped trajectory of active capacity over time. This curve is parameterized by two key elements:

\begin{itemize} 

\item \textbf{Final active capacity} $K$: This parameter represents the long-term, steady-state level of maintained technological capacity. It represents the ultimate level of equipment deployed and maintained for this particular technology.

\item \textbf{Characteristic deployment time} $\tau_{dep}$: this parameter reflects the timescale associated with achieving the final level of active capacity, $K$. It essentially captures the speed of deployment for the technology.

\end{itemize}

The model uses a weibull distribution to represent the variation in equipment lifespans before reaching their EoL. This distribution is characterized by two parameters:

\begin{itemize}

\item \textbf{Average equipment lifespan} $\tau_{EoL}$: this parameter represents the average time an equipment unit remains active before requiring replacement. It reflects the typical duration of service for individual equipment items.

\item \textbf{Coefficient of variation} $CV_{EoL}$: This parameter quantifies the degree of spread around the average lifespan ($\tau_{EoL}$). A higher $CV_{EoL}$ indicates greater variability in equipment lifespans. In simpler terms, it reflects the level of dispersion in how long individual equipment units last before needing replacement.
\end{itemize}

\begin{table}[h]
\centering
\resizebox{\linewidth}{!}{
\begin{tabular}{|c|c|c|c|}
 \hline
 Model inputs & Assumptions & Function\par chosen & Parameters \\
 \hline
 \hline
 Active\par capacity & \textbf{S-shaped} & logistic & $K$,\hspace{5pt}$\tau_{dep}$\\
 \hline
 EoL \par distribution & \textbf{Bell-shaped}& Weibull  & $\tau_{EoL}$,\hspace{5pt} $CV_{EoL}$\\
 \hline
\end{tabular}
}
\caption{Modeling assumptions and parameters. As shown in Section \ref{sec:AnalysisKeyParam} the results are robust to different choices of S-shaped and Bell-shaped curves, other than logistic and Weibull.}

\label{tab:param}
\end{table}

\subsection*{Results highlighted: industrial deployment dynamics}

Within the framework outlined and based on the adopted assumptions, the model predicts two qualitatively distinct behaviors for equipment production arising from the interplay between deployment speed and average equipment lifespan. These behaviors are summarized in Table~\ref{tab:res} for clarity.

\begin{enumerate}

\item \textbf{Fast deployment scenario:}

This scenario occurs when deployment is swift relative to the average equipment lifespan. Specifically, more than 60\% to 70\% of the final active capacity becomes operational within a single average equipment lifespan ($\tau_{EoL}$). During this rapid deployment phase, equipment production surpasses its renewal steady level, resulting in an initial overshoot. This overshoot is subsequently followed by damped oscillations around the steady-state production value.

\item \textbf{Slow deployment scenario:}

This scenario arises when deployment unfolds at a slower pace. In this case, less than 60\% to 70\% of the final capacity is deployed within a single average equipment lifespan. Consequently, equipment production exhibits a monotonic rise until it reaches its steady-state value associated with long-term renewal.
\end{enumerate}

\begin{table*}[h]
    \centering
    \resizebox{\linewidth}{!}{
        \begin{tabular}{|c||c|c|c|}
        \hline
        Behavior & Slow deployment & \multicolumn{2}{|c|}{Fast deployment} \\
        \hline
        \hline
        $\tau_{dep}$ & \textbf{Long} & \multicolumn{2}{|c|}{\textbf{Short}}\\
        \hline
        $\tau_{EoL}$ & \textbf{Short} & \multicolumn{2}{|c|}{\textbf{Long}}\\
        \hline
        $CV_{EoL}$ & High or Low & \textbf{High} & \textbf{Low} \\
        \hline
        \hline
        $\tau_{dep}$ relatively \par to $\tau_{EoL}$ & $\tau_{dep}>r_{c}\tau_{EoL}$ & \multicolumn{2}{|c|}{$\tau_{dep}<r_{c}\tau_{EoL}$}\\
        \hline
        \hline
        Time to reach target capacity & \textbf{Long} & \multicolumn{2}{|c|}{\textbf{Short}}\\ 
        \hline
        Overshoot & \textbf{No} & \multicolumn{2}{|c|}{\textbf{Yes}}\\ 
        \hline
        Oscillations & \textbf{No} & \textbf{Yes} & \textbf{No}, quickly damped\\
        \hline
        Transition\par time & \textbf{Short} & \textbf{Long} & \textbf{Short}\\
        \hline
        \end{tabular}
    }
\caption{Fast or slow characterization of deployment dynamics: Influence of parameters on production behavior.}
\label{tab:res}
\end{table*}

\subsection*{Mitigating industrial deployment strategies and business cycles}
\label{sec:KeyTakeaways}

First, the renewal of technological equipment in order to maintain a stable level of adoption of this technology depends on at least two factors: the speed at which this technology has been deployed, and the average lifespan of a piece of equipment.

This study identified a threshold that raises major obstacles and could entail critical damages to the industry. If deployment is slower than this threshold, the need for renewal will be smoothed out, and production will increase monotonically up to renewal steady production. If deployment is faster then renewal needs will come in waves, and production will oscillate. \textbf{This threshold has been found to be a deployment of between 60\% and 70\% of the ultimate capacity during one equipment lifespan.} 

Oscillation dynamics raise questions of production capacity sizing, which have repercussions on all sub-systems, whether material (number of plants, raw material requirements, energy supply system) or non-material (workforce, training path, regulation). Being able to cope with production peaks means operating below capacity during troughs. This low capacity utilization periods question the viability of the production actor, who risks to become stranded assets, and could therefore slow down investment. Employments specific to this technology will also fluctuate generating unemployment period, which can lead to skill maintaining issues. Consequently, it is crucial to consider the possibility of such a dynamic and adapt industrial deployment strategies accordingly.

At first glance, the potential actions of the industry seem to be of different kinds: to mitigate, bypass or adapt to this oscillating production dynamic. Reducing equipment service lifespan, lowering deployment speed or broadening EoL distribution can all help mitigate oscillations. However, these solutions raise questions of consumption and waste, in the case of the former, and of feasibility, in the case of the latter two. In a regionalized approach, an import/export policy can help bypass these oscillations. For example, size capacity to the production peak and then export, or size capacity to the renewal steady production and import to manage the peak. Finally, the industry can adapt to such oscillations by, for example, overseeing the transfer of skills from one generation to the next, or by cooperating with other players, notably the public sector, to adapt investment flows, subsidies and regulation.

\subsection*{Future development}

The proposed model provides an analytical framework that can be adapted to evaluate a diverse array of current and future technology deployment scenarios. Its versatility allows for the quantitative assessment of long-term renewal strategies across various technological domains, offering valuable insights for policy-makers and industry stakeholders.

While the current iteration of the model has demonstrated its utility, several possibilities for future research and refinement have been identified. These include incorporating multi-wave deployment modeling to capture more complex market dynamics, including evolving average lifetimes of technologies as they mature and improve, implementing parameterization to model incremental innovation through gradual reductions in material and energy requirements, expanding the model to encompass interdependent technology networks that collectively fulfill a specific service or function, and integrating economic variables and market forces to provide a more comprehensive, business-oriented perspective. 

The energy transition sector presents an interesting application for this model and its future iterations. The rapid deployment of renewable energy technologies such as wind turbines, solar panels, and electric vehicles, coupled with their extended operational lifespans, creates a complex landscape that necessitates sophisticated long-term planning tools. Extension of this model could contribute to decision-making in industrial deployment. Further empirical validation and sensitivity analyses should reinforce the model reliability and applicability across different technological and geographical contexts. Additionally, collaborations with industry partners and policy-makers would ensure the model relevance and practical utility in addressing deployment and renewal strategies.

\section*{Funding}

This work has benefited from a government grant managed by the Agence Nationale de la Recherche under the France 2030 program, reference ANR-22-PERE-0003.

\section*{Declaration regarding generative AI}
We hereby confirm that we have not used generative AI in the entire
process, neither for writing this manuscript nor for generating and
analysing data.

\section*{CRediT authorship contribution statement}
\textbf{Joseph Le Bihan:} Investigation, Methodology, Formal analysis, Conceptualization, Visualization, Writing - Original Draft. 

\textbf{Thomas Lapi:} Conceptualization, Methodology, Writing - Review \& Editing.

\textbf{José Halloy:} Conceptualization, Methodology, Writing - Original Draft, Writing - Review \& Editing, Supervision, Project administration, Funding acquisition.

\section*{Declaration of competing interest}
The authors affirm that no conflict of interest exists concerning the
present study.

\section*{Data access}
Data will be made available on request.


 \bibliographystyle{elsarticle-num} 
 \bibliography{references}





\end{document}